# Out-of-equilibrium criticalities in graphene superlattices

Alexey I. Berdyugin[1,2*,+], Na Xin[1,2+], Haoyang Gao[3], Sergey Slizovskiy[1,2], Zhiyu Dong[3], Shubhadeep Bhattacharjee[1,2], P. Kumaravadivel[1,2], Shuigang Xu[1,2], L. A. Ponomarenko[1,4], Matthew Holwill[1,2], D. A. Bandurin[1,2], Minsoo Kim[1,2], Yang Cao[1,2], M. T. Greenaway[5,6], K. S. Novoselov[2], I. V. Grigorieva[1], K. Watanabe[7], T. Taniguchi[8], V. I. Fal'ko[1,2,9], L. S. Levitov[3], R. Krishna Kumar[1,2,10*], A. K. Geim[1,2*]

[1]School of Physics & Astronomy, University of Manchester, Manchester M13 9PL, United Kingdom
[2]National Graphene Institute, University of Manchester, Manchester M13 9PL, United Kingdom
[3]Massachusetts Institute of Technology, Cambridge, Massachusetts 02139, USA
[4]Department of Physics, University of Lancaster, Lancaster LA1 4YW, United Kingdom
[5]Department of Physics, Loughborough University, Loughborough LE11 3TU, United Kingdom
[6]School of Physics & Astronomy, University of Nottingham, Nottingham NG7 2RD, United Kingdom
[7]Research Center for Functional Materials, National Institute for Materials Science, 1-1 Namiki, Tsukuba 305-0044, Japan
[8]International Center for Materials Nanoarchitectonics, National Institute for Materials Science, 1-1 Namiki, Tsukuba 305-0044, Japan
[9]Henry Royce Institute for Advanced Materials, Manchester M13 9PL, United Kingdom
[10]ICFO-Institut de Ciencies Fotoniques, The Barcelona Institute of Science and Technology, 08860 Castelldefels (Barcelona), Spain

+ These authors contributed equally to this work.
*Corresponding author. Email: alexey.berdyugin@manchester.ac.uk (A.I.B); roshankrishnakumar90@gmail.com (R.K.K.); geim@manchester.ac.uk (A.K.G.)

**In thermodynamic equilibrium, current in metallic systems is carried by electronic states near the Fermi energy whereas the filled bands underneath contribute little to conduction. Here we describe a very different regime in which carrier distribution in graphene and its superlattices is shifted so far from equilibrium that the filled bands start playing an essential role, leading to a critical-current behavior. The criticalities develop upon the velocity of electron flow reaching the Fermi velocity. Key signatures of the out-of-equilibrium state are current-voltage characteristics resembling those of superconductors, sharp peaks in differential resistance, sign reversal of the Hall effect, and a marked anomaly caused by the Schwinger-like production of hot electron-hole plasma. The observed behavior is expected to be common to all graphene-based superlattices.**

The electric response of metallic systems is routinely described by a Fermi surface displacement in momentum space, established through a balance between acceleration of charge carriers and their relaxation caused by scattering (*1*). The displacement is usually small, such that the drift velocity $v_d$ is minute compared to the Fermi velocity $v_F$. In theory, if inelastic scattering is sufficiently weak, it should be possible to shift the Fermi surface so far from equilibrium that all charge carriers within the topmost, partially filled bands start streaming along the applied electric field $E$. The field would then start producing extra carriers via interband transitions (*2*), allowing electronic bands under the Fermi energy to contribute to the charge flow. Such an extreme out-of-equilibrium regime has never been achieved in metallic systems because Ohmic heating, phonon emission and other mechanisms greatly limit $v_d$ (*3–5*).

A rare exception is semi-metallic graphene. At high carrier densities $n$, the drift velocity in graphene is limited by phonon emission (*6, 7*), similar to other metallic systems. However, at low $n$, thermal excitations can create a relativistic plasma of massless electrons and holes, the 'Dirac fluid'. Its properties in thermodynamic equilibrium were in the focus of recent research (*8–12*) but the behavior at high biases represents an uncharted territory. Yet, close to the Dirac point even a small $E$ can shift the entire Fermi surface and tap into a supply of carriers from another band (*13*, *14*). This can trigger processes analogous to the vacuum breakdown and Schwinger particle-antiparticle production in quantum electrodynamics where they are predicted to occur at enormous fields of ~$10^{18}$ V m$^{-1}$ (*15*). Because such *E* are inaccessible, it is enticing to mimic the Schwinger effect and access the resulting out-of-equilibrium plasma in a condensed matter experiment (*13*, *14*, *16*). Certain nonlinearities observed near graphene's neutrality point (NP) were previously attributed to the creation of electron-hole (e-h) pairs by a Schwinger-like mechanism (*13*, *14*) but the expected intrinsic behavior was obscured by low mobility, charge inhomogeneity and self-gating effects (*6*, *17*).

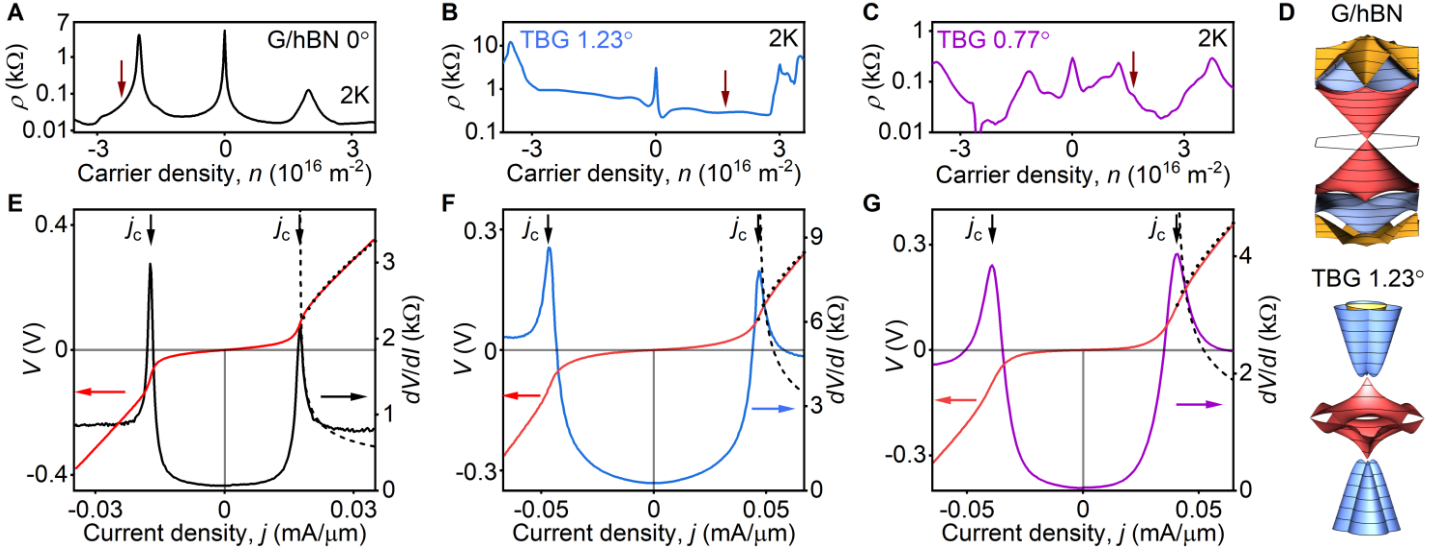

**Fig. 1. Linear and nonlinear transport in graphene superlattices.** (**A-C**) $\rho(n)$ *in the linear regime (j = 50 nA/μm) for G/hBN with θ ≈ 0°(A), and TBG with θ ≈ 1.23°(B) and θ ≈ 0.77°(C). Micrographs of the studied devices are provided in (29).* (**D**) *Band structures of G/hBN and TBG 1.23° superlattices (see 29). Colors denote different energy bands. The bands are shown for the energy range of ±340 and ±80 meV for G/hBN and TBG 1.23°, respectively.* (**E-G**) *IV characteristics for the devices in panels (A-C), respectively. The doping levels for the curves are marked by the arrows in A-C. The dependence (j- $j_c$) $\propto V^{3/2}$ expected above $j_c$ is shown by the dotted curves and the corresponding dV/dI $\propto$ (j- $j_c$)$^{-1/3}$ by the dashed curves. All V and dV/dI are normalized according to devices' aspect ratios.*

Here we use graphene-based superlattices to identify an out-of-equilibrium state that sharply develops above a well-defined critical current $j_c$. The current marks an onset of the Schwinger pair production and a transition from a weakly-dissipative fluid-like flow to a strongly dissipative e-h plasma regime. The out-of-equilibrium Dirac fluid is realized at surprisingly small *E*, thanks to the narrow electronic bands and low $v_F$ characteristic of graphene superlattices (*18*, *19*). The resulting dual-band transport leads to striking anomalies in longitudinal and Hall resistivities. Counterintuitively, an apparent drift velocity in this regime exceeds $v_F$. With hindsight, we show that the current-induced critical state can be reached even in standard graphene, using extra-high currents allowed by the point contact geometry.

The studied superlattices were of two types: graphene crystallographically aligned on top of hexagonal boron nitride (G/hBN) (*20–23*) and small-angle twisted bilayer graphene (TBG) (*24–28*). The superlattices were encapsulated in hBN, to ensure high electronic quality, and shaped into multiterminal Hall bar devices using the standard fabrication procedures (*29*). The devices were first characterized by measuring their longitudinal resistivity $\rho$ as a function of *n* as shown in Figs. 1, A-C, for three representative devices. The twist angles $\theta$ were determined from measurements of Brown-Zak oscillations (*30*); for TBG, $\theta$ was intentionally chosen away from the magic angle to avoid many-body states (*27*, *28*). Aside from the familiar peak in $\rho$ at zero doping, satellite peaks indicating secondary NPs were observed at *n* that agreed well with the $\theta$ values (*20–22*, *26*). For G/hBN superlattices, the low-energy electronic spectrum is practically identical to that of monolayer graphene (*18*), and the spectral reconstruction occurs only near and above the edge of the first miniband (Fig. 1D, top). In contrast, all minibands in TBG are strongly reconstructed (*19*) (Fig. 1D, bottom). At low biases (Figs. 1, A-C; *fig. S1*), our devices exhibited transport characteristics similar to those reported previously for G/hBN and TBG superlattices (*20–22, 26*).

Next, we studied high-bias transport using current densities $j$ up to 0.1 mA/μm, limited only to avoid device damage. Unless stated otherwise, all the reported measurements were carried out at the bath temperature $T =$ 2 K. The superlattices exhibited qualitatively similar current-voltage (IV) characteristics (Figs. 1, E-G), which were nearly linear at $j < 0.01$ mA/μm and then rapidly switched into a high-resistance state so that the differential resistivity $dV/dI$ showed a pronounced peak at a certain critical current $j_c$. The behavior was universal, found in all our devices (>10) (see *figs. S3, S6*), if the Fermi energy was tuned inside narrow minibands (that is, away from the main NP in the case of G/hBN). The *I-V* characteristics in Figs. 1, E-G, strikingly resemble the superconducting response, despite electron transport being ballistic at low $j$ and viscous at moderate currents (*31*); $\rho$ always remained finite although could be as low as <0.01 kΩ, a few orders of magnitude smaller than $dV/dI$ above $j_c$. Figure 2 provides further details by showing $dV/dI$ as a function of $n$, where the narrow white arcs correspond to peaks in $dV/dI$. Considerable similarities are clearly seen across different superlattice types. One interesting feature shared by all the maps was the rapidly decreasing $j_c$ as $n$ approached NPs (Figs. 2, A-C; *figs. S2, S6*). The only exception was the main NP in G/hBN superlattices where the resistivity in its vicinity increased monotonically for all accessible $j$ (*fig. S2*).

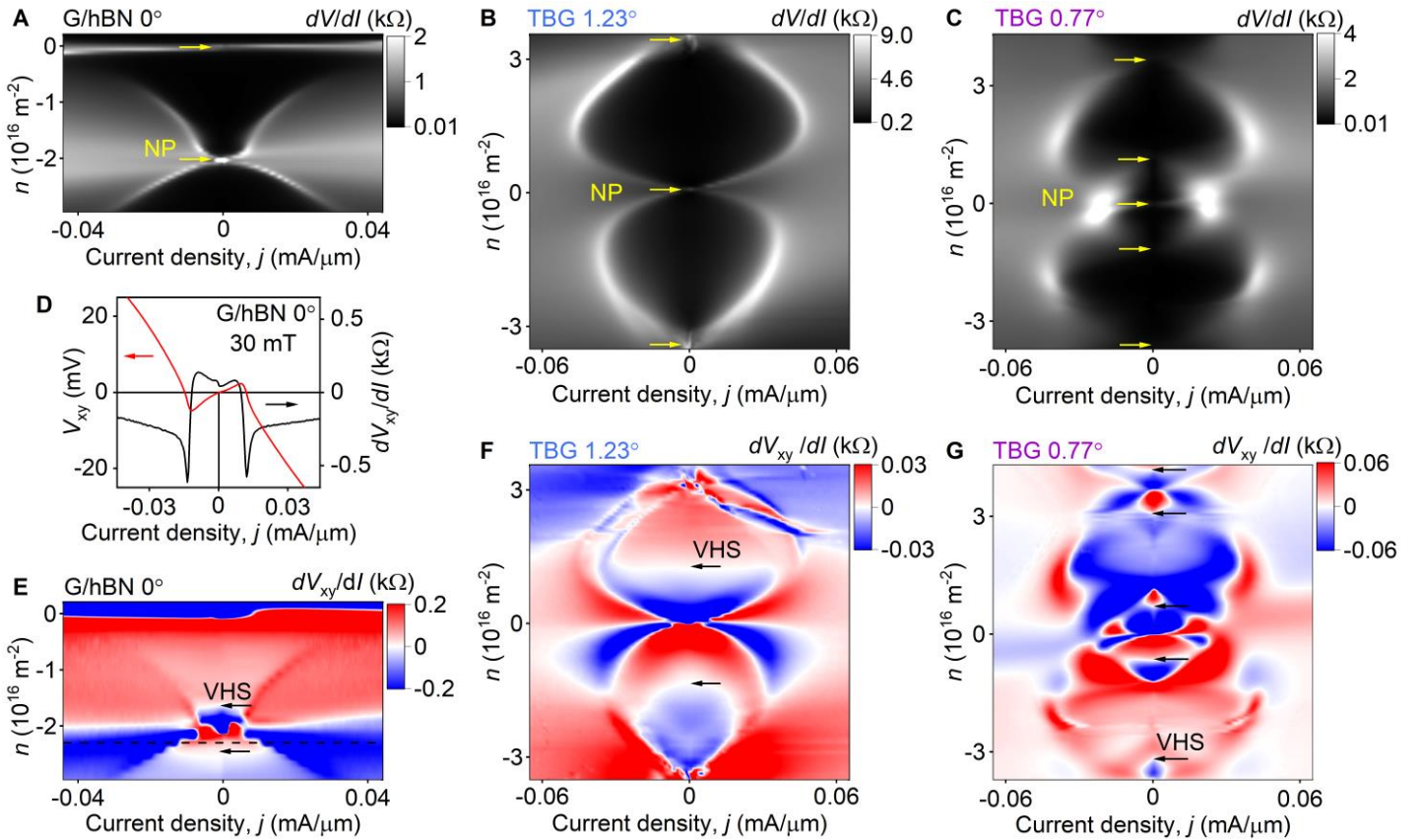

**Fig. 2. Switching into the high-bias regime.** (**A-C**) *$dV/dI$ as a function of $j$ and $n$ for the superlattices in Figs. 1, A-C, respectively. Bright arcs signify the critical current. Yellow arrows: NPs as found from low-bias measurements (29).* (**D**) *Hall voltage (red curve) and the corresponding differential resistivity (black) measured at n indicated by the dashed line in (E).* (**E-G**) *Maps of $dV_{xy}/dI$ for the superlattices in (A-C), respectively. B = 30 mT; T = 2 K. The black arrows mark positions of van Hove singularities.*

To gain more insight, we studied the Hall effect in small (non-quantizing) magnetic fields $B$. Figure 2D shows an example of such measurements for G/hBN near the hole-side NP. At small $j$, the Hall voltage $V_{xy}$ increased linearly with $j$ and $dV_{xy}/dI$ was positive, reflecting the hole doping. However, $dV_{xy}/dI$ abruptly turned negative above $j_c$, revealing a change in the dominant-carrier type. Figures 2, E-G, show $dV_{xy}/dI$ maps for the G/hBN and TBG superlattices. One can see clear correlations between the longitudinal and Hall maps such that the peaks in $dV/dI$ and the Hall effect's reversal occurred at same $j_c$. The observed nonlinearities were robust against $T$ up to ~50 K, above which the peaks in $dV/dI$ became gradually smeared (*fig. S4*). This shows that Ohmic heating – generally expected at high $j$ (*14, 31, 32*) – was not the reason for the critical-current behavior (*29*).

The rapid decrease in $j_c$ near all secondary NPs prompts the question why such a critical-current behavior was not observed in graphene (*13, 14*) or near the main NP of G/hBN (Fig. 2A) and whether it can be achieved at some higher $j$. With this in mind, we employed a point contact geometry that funneled the current through a short constriction whereas wide adjacent regions provided a thermal bath for electron cooling. This allowed us to reach $j$ an order of magnitude higher than those achievable in the standard geometry. At these $j$, IV characteristics near the main NP of G/hBN superlattices became similar to those near its secondary NPs (*fig. S3*), although they were more smeared because of Ohmic heating and, possibly, edge irregularities in the superlattice periodicity within narrow constrictions. To circumvent the latter problems and demonstrate the universality of the critical behavior at all NPs, we made constrictions from non-superlatticed graphene (monolayer graphene encapsulated in hBN but nonaligned). These devices also displayed a clear critical behavior although peaks at $j_c$ were notably broader because of heating (Fig. 3A).

To understand the criticalities, let us first discuss the conceptually simplest case of the Dirac spectrum, as in non-superlatticed graphene. We consistently observed that the transition between the low and high resistance states occurred at $j_c \approx nev_F$ ($e$ is the electron charge), that is, at $v_d \approx v_F$, independently of $n$ (Fig. 3B). This condition means that the Fermi surface is shifted from equilibrium by the entire Fermi momentum and, as illustrated in Fig. 3C, all electrons in the conduction band move along $E$ with a drift velocity of about $v_F$. If the spectrum were fully gapped, $j$ could not increase any further because all available carriers already move at maximum speed. This should result in saturation of $j$ as a function of $V$, in agreement with the observations at $j \lesssim j_c$. Simulations of this intraband-only transport corroborate our conclusions (Fig. 3A, dashed curves). To explain the supercritical behavior at $j > j_c$, we note that, for a gapless spectrum, $E$ can move electrons up in energy from the valence band into the conduction band, leaving empty states (holes) behind (Fig. 3C, bottom panel). The extra electrons and holes created by the interband transitions allow the current to exceed $j_c$. Accordingly, the apparent $v_d = j/ne$ seemingly exceeds the maximum possible group velocity, $v_F$ (because $n$ is fixed by gate voltage, but the actual concentration of carriers increases by $\Delta n$). Quantitatively, the e-h production at $j > j_c$ can be described by the Schwinger (or Zener-Klein tunneling) mechanism. It can generate interband carriers at a rate $\propto E^{3/2}$ (*13, 16*) but at small biases the production is forbidden by the Pauli exclusion principle. Above $j_c$ the Fermi distribution is shifted sufficiently far from equilibrium so that $E$ depletes the states near the NP, which eliminates the Pauli blocking and enable the e-h pair production (Fig. 3C). Accounting for e-h annihilation (recombination processes bring the electronic system back into the equilibrium), we find the stationary concentration of extra carriers $\Delta n$ to be $\propto E^{3/2} \propto V^{3/2}$, if $\Delta n \ll n$ (*29*). This translates into extra current $\Delta nev_F \propto V^{3/2}$ and $dV/dI \propto j^{-1/3}$. As $dV/dI$ decreases for $j > j_c$ but increases for $j < j_c$, a peak is expected at $j_c$, in agreement with Fig. 3A.

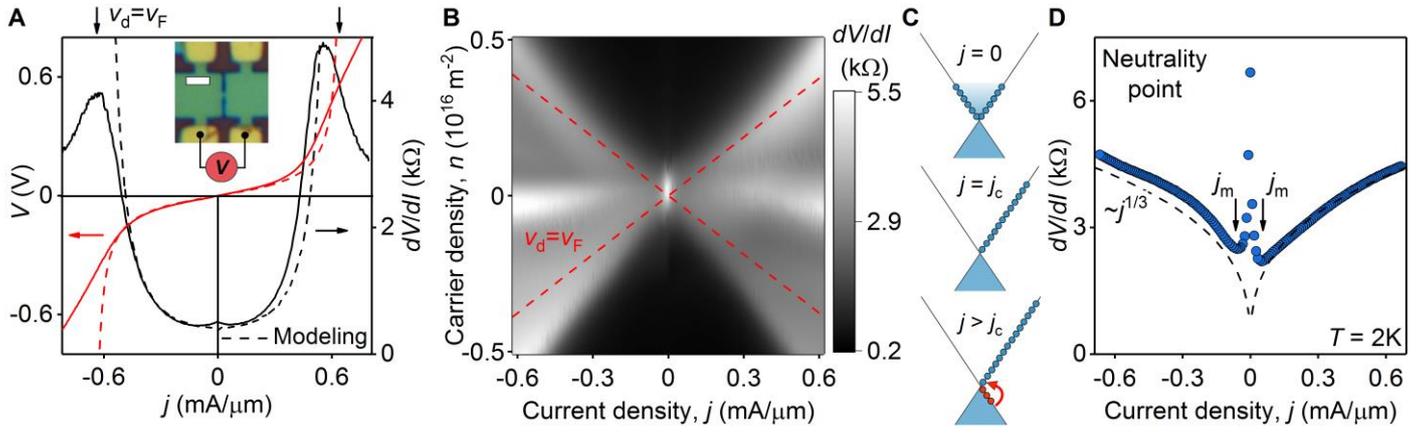

**Fig. 3. Nonlinear transport in non-superlatticed graphene near the Dirac point.** (**A**) *Voltage and differential resistance (red and black curves, respectively) for a constriction of 0.4 µm in width; n = 0.4×10$^{16}$ m$^{-2}$. Inset: Optical micrograph of the graphene device and its measurement geometry. Scale bar, 2 µm. The small bump at zero bias is caused by electron-electron scattering (34). Dashed curves: IV characteristics calculated for the Dirac spectrum at j < j$_c$ (29). The vertical arrows indicate j with v$_d$ = v$_F$ = 1×10$^6$ m s$^{-1}$.* (**B**) *Example of dV/dI maps for graphene constrictions. Red lines: j = nev$_F$.* (**C**) *Schematic of graphene's spectrum and its occupancy in equilibrium (top) and out-of-equilibrium for j = j$_c$ = nev$_F$ (middle) and j > j$_c$ (bottom). Blue and red circles: electrons and holes, respectively. The red arrow illustrates e-h pair production.* (**D**) *dV/dI at the NP for a 0.6 µm wide constriction. The arrows mark minima.*

The above analysis can also be applied to graphene superlattices. Their narrow minibands display low $v_F$ and, therefore, the onset of interband transitions is expected at small $j$. Indeed, the switching transition in our superlattices occurred at $v_d$ typically >10 times smaller than in non-superlatticed graphene (*fig. S5*). This yields a characteristic $v_F$ of several 10$^4$ m s$^{-1}$, which translates into minibands' widths of ~10 meV, as expected from band structure calculations (*19*). For the relatively small $j_c$, superlattices were much less affected by heating than graphene and, accordingly, exhibited sharper transitions (cf. Figs. 2 & 3A). Figures 1, E-G, compare the experimental IV curves with the above predictions for Schwinger-like carrier generation. Good agreement is found for $j \gtrsim j_c$. Notable deviations seen at highest $j$ are expected because $\Delta n$ is no longer small compared to $n$, the assumption used to derive the plotted dependences (*29*). Furthermore, $j_c$ in graphene evolved $\propto n$ as expected for the Dirac spectrum (Fig. 3B). In contrast, superlattices exhibited clear deviations from the linear dependence (Figs. 2, A-C). This is attributed to the fact that the group velocity of charge carriers rapidly decreases away from secondary NPs, dropping to zero at van Hove singularities (VHS). If nonequilibrium carriers reside near VHS, they move at low speed and contribute little to the current (*fig. S5C*), leading to the sublinear $j_c(n)$ as observed experimentally.

Extending the described physics onto the Hall effect, it is straightforward to understand the sign changes in Figs. 2, D-G. With reference to Fig. 3C, interband transitions result in extra holes near the NP plus extra electrons that effectively appear at higher energies in the out-of-equilibrium Fermi distribution (Fig. 3C). For superlattices, contributions of these e-h pairs into $V_{xy}$ do not cancel each other because of the broken e-h symmetry, which results in different masses and mobilities of the extra carriers. The effect is particularly strong on approaching VHS. For example, if the dominant carriers are electrons, their distribution would be shifted by $E$ upwards towards a VHS (*fig. S5C*), and they should have heavy masses. In contrast, the reciprocal holes generated near the NP should be light (*fig. S5C*). These higher-mobility holes are expected to provide a dominant contribution into the Hall signal and, therefore, $dV_{xy}/dI$ should change its sign from electron to hole near $j \approx j_c$, as observed experimentally. If the asymmetry is sufficiently strong, even $V_{xy}$ can reverse its sign (Fig. 2D). The observed changes in the Hall effect can qualitatively be described using the two-carrier model with different mobilities of out-of-equilibrium electrons and holes (*fig. S7*).

Finally, we discuss the interband carrier generation at the main NP in graphene (Fig. 3), which closely mimics the Schwinger effect in quantum electrodynamics. Consequences of the Schwinger-like effect at the Dirac point are qualitatively different from those described by Zener-Klein tunneling at finite doping (*29*). Figure 3D shows that, in contrast to the latter case, there is no low-to-high resistance switching at $n = 0$, and $dV/dI$ rapidly drops with increasing $j$, reaches a minimum and then gradually increases. This behavior was highly reproducible for all graphene constrictions (*fig. S8*) but, because of self-gating and heating effects, could not be observed in the standard geometry where IV curves were similar to those in the literature (*6*). The initial drop is attributed to e-h puddles present at NPs wherein small $E$ starts generating interband carriers along puddles' boundaries and enhances conductivity (*13*). Indeed, minima in $dV/dI$ typically occurred at $j_m \approx 0.05$ mA µm$^{-1}$ (Fig. 3D) which translates into $\Delta n = j_m/ev_F \approx 3\times 10^{10}$ cm$^{-2}$, in agreement with the charge inhomogeneity $\delta$ found in our devices. In principle, the initial $dV/dI$ drop could be fitted again by $\propto j^{-1/3}$ but such fits were inconclusive because of the involved inhomogeneity. For higher $j$ such that $\Delta n \gg \delta$, the Schwinger production fills graphene with a plasma of electrons and holes in equal concentrations, $n_e \approx n_h = \Delta n$. Because the annihilation rate of e-h pairs scales with $n_e n_h = \Delta n^2$, theory predicts (*29*) that the Schwinger production rate ($\propto E^{3/2}$) leads to $\Delta n \propto E^{3/4}$, resulting in $dV/dI \propto j^{+1/3}$. This contrasts the reported Zener-Klein behavior at graphene's NP (*13*) but is in quantitative agreement with our experiment (Fig. 3D, *fig. S8*). For highest $j$, the hot e-h plasma inside graphene constrictions is expected to approach the quantum critical limit (*8–12*), in which e-h scattering is governed by the uncertainty principle and $\rho$ is predicted to become rather universal, $\sim 1.3\alpha^2(h/e^2)$ where $\alpha$ is the interaction constant and $h/e^2$ the resistance quantum (*8, 9*). For encapsulated graphene, $\alpha \approx 0.3$ whereas the constriction geometry results in resistance of $\sim 1.8\rho$ (*29*). Accordingly, the quantum-critical resistance for our constrictions is expected to be ~5 kΩ, in a qualitative agreement with Fig. 3D and *fig. S8* where the curves approach this value. We do not expect better agreement because $E$ strongly disturbs the electron-hole plasma making it anisotropic, rather different from the Dirac fluids in thermal equilibrium, which were discussed previously (*8–12*). This anisotropic regime requires further theoretical analysis and would be interesting to probe by other experimental techniques.

To conclude, at high biases, Fermi liquids in graphene-based systems can be turned into Dirac-like fluids characterized by intense interband carrier generation. The transition between the weakly and strongly dissipative electronic states is marked by peculiar superconducting-like $dV/dI$. Such IV characteristics, while of interest on their own right as a signature of out-of-equilibrium criticalities, also serve as a warning that they alone - without other essential attributes (e.g., zero resistance) – do not constitute a proof of 'emerging/fragile' superconductivity. It is possible that the nonlinear response reported in some graphene-based flat-band systems (e.g., (*33*)) was governed by the out-of-equilibrium physics rather than superconductivity. Other attributes of nonequilibrium behavior such as Bloch oscillations and associated THz radiation are likely to accompany the reported criticalities, an appealing opportunity for further investigation.

**Supplementary Information.**

#1 Device fabrication and electrical measurements

The van der Waals (vdW) heterostructures studied in our work were assembled by the dry transfer method (*36*, *37*) using stamps made from poly-bisphenol A carbonate (PC) and polydimethylsiloxane (PDMS). To fabricate G/hBN superlattices, first an hBN crystal (30-80 nm thick) was picked up by such a polymer stamp. The crystal would eventually serve as the top hBN layer of the heterostructure. This was followed by picking up monolayer graphene and then the bottom hBN layer (20-50 nm thick). During the trilayer assembly the substrate temperature was kept at 80-90 °C. Graphene and hBN crystals chosen for the vdW assembly had straight edges so that their crystallographic axes could be aligned (*20*). Only one of the hBN crystals (either top or bottom) was aligned while the other was intentionally misaligned by ~15° with respect to graphene axes. Both hBN crystals were misaligned in the case of non-superlatticed graphene. The trilayer stacks were released on top of an oxidized Si wafer (300 nm of $SiO_2$) at a temperature of 170-180 °C. The resulting vdW heterostructure were checked by Raman spectroscopy to verify the presence (or absence) of crystallographic alignment between graphene and hBN (*38*, *39*).
Similar stacking procedures were used to create TBG superlattices but, instead of monolayer graphene, twisted bilayer graphene was encapsulated between top and bottom hBN. Twisted bilayers were prepared by the cut and stack technique (*40*). To this end, a single crystal of monolayer graphene was cut in halves using a sharp tip of an atomic force microscope. After picking up one of the halves, the assembly stage was rotated by ~1º and the second half was subsequently picked up onto the same PC/PDMS stamp.
After the described vdW heterostructures were finally placed on top of a Si wafer, we defined contacts regions using electron beam lithography and reactive ion etching. Cr/Au (3 nm/60 nm) films were deposited into the etched regions to provide quasi-1D electrical contacts to graphene (*36*). As the final step, the same lithography and etching procedures were employed to shape the vdW assembly into devices having either Hall bar or constriction geometry (figs. S1, A-C).
To vary doping in our devices, we applied DC voltage between graphene and the silicon wafer. The standard low-frequency lock-in technique was employed to characterize the devices in the linear response regime using small excitation currents of 0.1-1 µA. To study high-bias response, we mixed AC and DC currents using the standard scheme of applying AC and DC voltages through large resistors. Both DC and AC voltages were measured using the 4-probe configurations whilst the applied DC current was monitored by a separate current meter.

#2 Hall effect at low biases
Figures S1, D-F show Hall resistivity $R_{xy}$ as a function of the carrier density $n$ induced by gate voltage for the three superlattice devices reported in the main text. In these and all the other studied superlattices, $R_{xy}$ changed its sign multiple times. The change occurred each time when the Fermi energy passed through inflection points in the electronic spectrum. Detailed behavior of $R_{xy}$ near inflection points depended on whether those were van Hove singularities (VHS) or neutrality points (NPs). For NPs, $R_{xy}$ switched its sign abruptly, tending to diverge at either side of the NP, which indicated a change from small electron to small hole doping. This divergence was accompanied by pronounced maxima in resistivity $\rho$ as seen in Figs. 1A-C of the main text. In contrast, $R_{xy}$ changed smoothly through zero at VHS, indicating a high density of states for both electrons and holes residing near VHS. Peaks in $\rho$ were either absent or minor at VHS. This qualitatively different behavior near NPs and VHS allowed us to identify their positions in all our devices as illustrated in figs. S1, D-F.

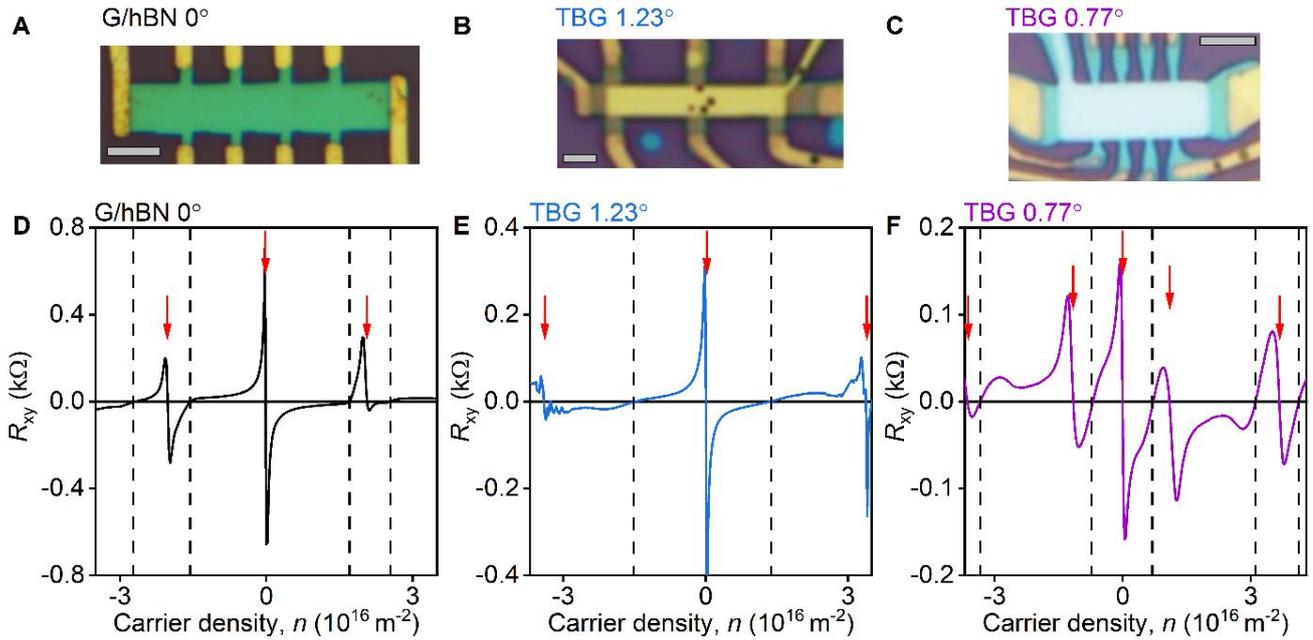

**Fig. S1. Hall Effect measurements in the linear response regime.** (**A-C**) *Optical micrographs of the superlattice devices used in Figs. 1 and 2 of the main text. (A) G/hBN superlattice. Scale bar, 4 μm. Contact resistances at 2 K: ~5 kΩ at zero back gate voltage and about 1 kΩ between NPs. (B) TBG 1.23°; scale bar, 1.5 μm. Contact resistances: 4-7 kΩ at the main NP and down to 1-2 kΩ at higher n. (C) TBG 0.77°; scale bar, 3.5 μm. Contact resistances of narrow contacts were 5-7 kΩ at zero back gate voltage.* (**D**) *$R_{xy}(n)$ for a G/hBN superlattice. The gate-to-graphene capacitance was determined from $R_{xy}$ measurements around the main NP. The red arrows and black lines mark NPs and VHS, respectively.* (**E-F**) *Same as in (D) but for TBG 1.23º and TBG 0.77º devices, respectively. T = 2 K; B = 30 mT for all the panels.*

#3 Critical-current behavior near all neutrality points in G/hBN superlattices
Figure 2A of the main text provided the differential resistivity map for hole doping of one of our G/hBN superlattice devices. This *dV/dI* map is extended into the electron doping regime in fig. S2A. While one can see a sharp transition from low to high dissipation transport for the case of hole doping, the map shows blank for electron doping. Nonetheless, if the map resolution is increased by a factor of 10, somewhat similar critical behavior could be resolved near the secondary NP in the conduction band, too (fig. S2B). In this figure, the transition between low- and high- resistance states occurred at approximately the same $j_c$ and $n$ as in the valence band, and the critical current also rapidly converged to zero as the Fermi energy approached the electron-side NP. This suggests that the critical behavior observed around both secondary NPs were governed by the same mechanism as discussed in the main text. The difference is that the transition for electron doping was much less sharp, so that no peak in *dV/dI* appeared at $j_c$ (fig. S2C). This smeared behavior can be attributed to the presence of another electron energy band in the spectrum, which overlaps with the secondary NP (see Fig. 1D of the main text). The extra band is expected to provide a parallel conduction channel, obscuring the transition from a viscous flow to e-h plasma transport. Fig. S2D also shows that no switching transition could be observed in our Hall-bar G/hBN devices near the main NP for all accessible *j* and, instead, *dV/dI* gradually increased with increasing *j*.

To check for possible critical-current behavior near the main NP at higher *j*, we made G/hBN devices in the point contact geometry, similar to that shown in the inset of Fig. 3A. The constrictions allowed us to reach an order of magnitude higher *j* and observe the low-to-high dissipation transition near the main NP, too (figs. S3 B,C). The required $j_c$ were close to those in non-superlatticed graphene (that is, $v_d \approx 1 \times 10^6$ m s$^{-1}$), as expected because spectra of G/hBN superlattices at low doping are practically identical to graphene's spectrum (Fig. 1D of the main text).

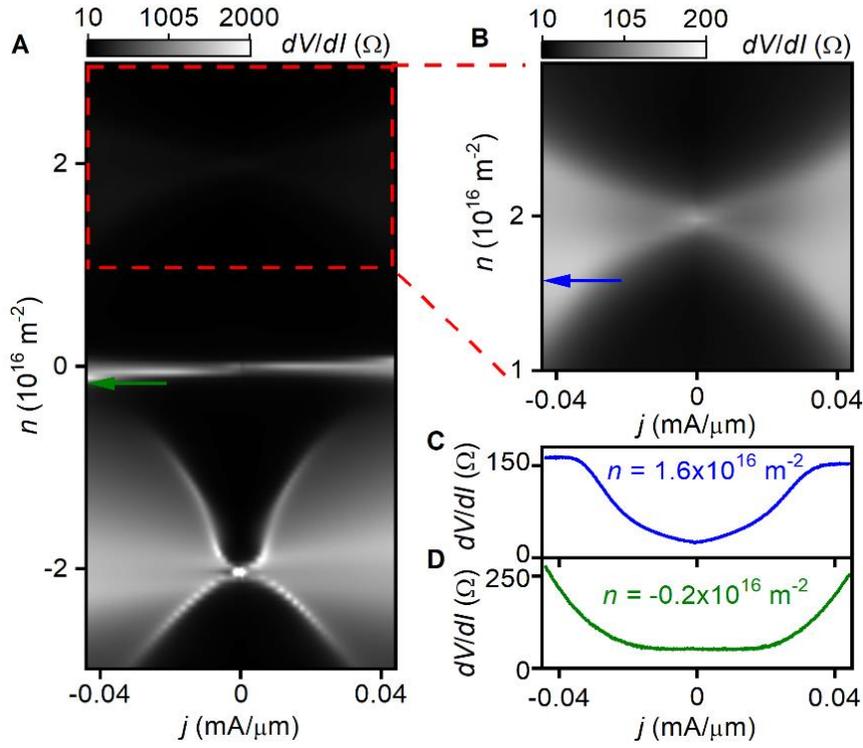

**Fig. S2. High bias transport in G/hBN superlattices.** (**A**) *Map of differential resistivity for the device in Fig. 2A of the main text, which now includes positive n (electron doping).* (**B**) *Zoom-in for the area indicated by the red box in (A).* (**C**) *Cross section of the map in (B) at the fixed electron density marked by the blue arrow.* (**D**) *Similar to (C) but for n near the main NP as indicated by the green arrow in (A).*

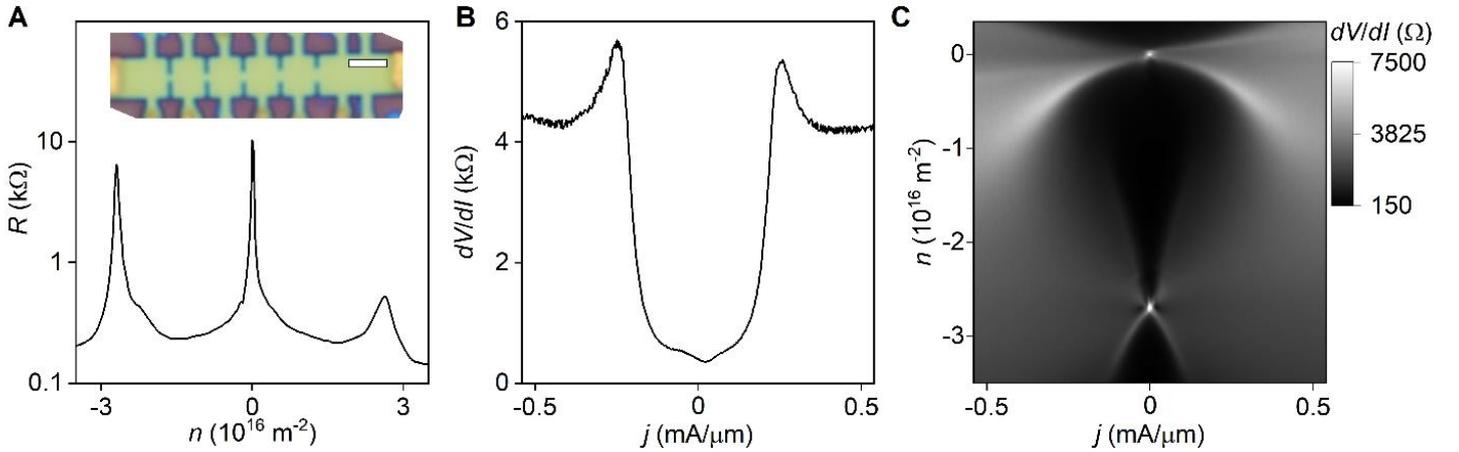

**Fig. S3. Critical-current behavior in G/hBN constrictions.** (**A**) *Resistance of a 0.4 μm wide constriction in the linear response regime. Inset: Optical micrograph of the G/hBN device with several constrictions. Scale bar, 2 μm. Characteristic contact resistance at zero gate voltage, 2-6 kΩ.* (**B**) *Differential resistance for the same device near the main NP; $n = -0.3 \times 10^{16}$ m$^{-2}$.* (**C**) *dV/dI maps for the same constriction. Note that the switching transition near the hole-side NP occurred at notably higher $j_c$ (~5 times) than those found in our Hall-bar devices made from same G/hBN superlattices (Fig. 2A, fig. S6). This seeming inconsistency is attributed to dominant contributions from wider contact regions on either side of the constrictions, where the switching transition near the secondary NP occurred earlier, at j lower than those calculated within the constriction itself. T = 2 K for all the panels.*

#4 Temperature dependence of the switching transition

The observed sharp transition from Fermi-liquid to e-h plasma transport was smeared with increasing $T$, but this happened well above liquid-helium temperatures. Examples of our measurements of $dV/dI$ at different $T$ are shown in fig. S4. One can see that the peaks in differential resistivity did not shift but became increasingly broader with increasing $T$.

For G/hBN, the sharp transition between low and high resistance states became completely invisible above liquid-nitrogen $T$ and, instead, $dV/dI$ gradually increased with $j$ (fig. S4A). On the other hand, constrictions made from non-superlatticed graphene exhibited pronounced peaks and step-like changes, even at $T$ close to room temperature (fig. S4B). As discussed in the main text and explained in further detail below, the peak in $dV/dI$ is expected only if the amount of generated interband carriers is sufficiently small ($\Delta n \ll n$). In this case, $dV/dI$ above but close to $j_c$ should decrease with increasing $j$ as $\propto j^{-1/3}$, in contrast to the two adjacent regimes of low $j < j_c$ and high $j \gg j_c$ where the differential resistivities are expected to increase with $j$. If the concentration $\Delta n$ of additional carriers created by $T$ is no longer small compared to $n$, e-h annihilation changes the $dV/dI$ dependence qualitatively. Therefore, the narrow region with $dV/dI \propto j^{-1/3}$ should vanish at high $T$, leading to disappearance of the peak in $dV/dI$, as observed experimentally and expected in theory (see the next section).

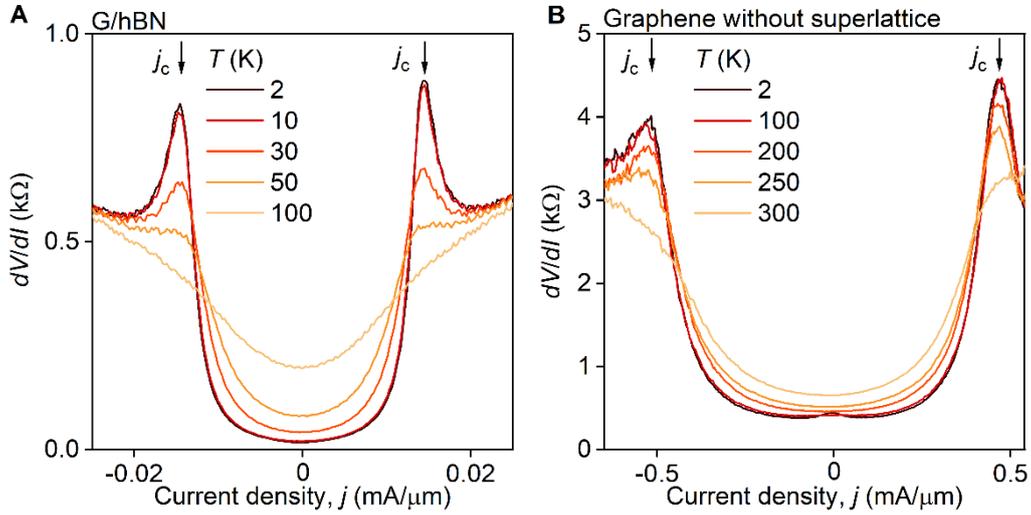

**Fig. S4. IV characteristics at higher temperatures.** (**A**) *Differential resistivity at different T measured for a G/hBN Hall bar device near the hole-side secondary NP; $n = -1.2 \times 10^{16}$ m$^{-2}$ and $\theta \approx 0°$.* (**B**) *Same as (A) but for a non-superlatticed graphene constriction with $W \approx 0.4$ μm; $n = 0.34 \times 10^{16}$ m$^{-2}$. The vertical black arrows indicate $j_c$.*

#5 Single band electron transport simulations in graphene

Below we explain our simulations of *I-V* characteristics for monolayer graphene at current densities $j < j_c$, which are shown by the dashed curves in Fig. 3A of the main text.

First, assuming that our system is in the diffusive regime, we calculate the drift velocity $v_d$ from $j$ and the carrier density $n$ induced by gate doping as

$$v_d = j/ne \qquad (S1)$$

Then, we determine the voltage drop produced by the current as a function of the drift velocity using the kinetic equation

$$\partial_t f + v_x \partial_x f + eE \partial_{p_x} f = St[f] + St_{e-e}[f] \qquad (S2)$$

$$St[f] = \int \frac{d^2 p_1}{(2\pi\hbar)^2} \left[ -f(\mathbf{p})(1 - f(\mathbf{p}_1))W_{\mathbf{p}\to\mathbf{p}_1} + f(\mathbf{p}_1)(1 - f(\mathbf{p}))W_{\mathbf{p}_1\to\mathbf{p}} \right] \tag{S3}$$

where $St[f](\mathbf{p})$ is the collision integral due to phonons and impurities, and $St_{e-e}$ is the collision integral due to e-e scattering. We integrate Eq. (S2) over the phase space with the weight $p_x$. Assuming the stationary and homogeneous solution ($\partial_t f = 0; \partial_x f = 0$) and using the fact that e-e collisions conserve the total momentum ($\int d^2 p\, p_x St_{e-e}[f](\mathbf{p}) = 0$) we get

$$eEn = \int \frac{d^2 p_1 d^2 p}{(2\pi\hbar)^4} (p_x - p_{1x}) f(\mathbf{p})(1 - f(\mathbf{p}_1))\, W_{\mathbf{p}\to\mathbf{p}_1} \tag{S4}$$

where $eE$ is the voltage drop per unit length.
To calculate the integral in eq. (S4) we take the scattering rate in the form $W_{\mathbf{p}\to\mathbf{p}_1} = C \cos[(\theta - \theta_1)/2]^2$, as dictated by the structure of graphene wavefunctions (41), where $C$ is the numerical constant that characterizes the number of scatterers inside the channel. Additionally, we assume that e-e interactions sufficiently equilibrate the carrier distribution so that carriers have the stationary thermal distribution in a frame moving with the velocity $v_d$ along the electric field

$$f(\mathbf{p}, x) = n_F(\epsilon_p - p_x v_d - E_F, T) = \frac{1}{e^{(\epsilon_p - p_x v_d - E_F)/k_B T} + 1} \tag{S5}$$

where $\epsilon_p = p v_F$, $v_F$ is the Fermi velocity in graphene and $k_B$ is the Boltzmann constant. To obtain $E_F$ for a moving frame we assume $E_F \gg T$ and normalize the distribution function (S5) using the initial doping condition

$$(\pi\hbar)^{-2} \int d^2 p\, n_F(\epsilon_p - p_x v_d - E_F, T) = n \tag{S6}$$

This leads to

$$E_F^2 = n\pi\hbar^2 v_F^2 (1 - v_d^2/v_F^2)^{3/2} \tag{S7}$$

From equation (S7) we find $E_F$. Next, we use distribution function (S5) to calculate integral (S4) for each current density value. We adjust constant $C$ in the scattering rate to match the measured resistivity around zero current and then use the same $C$ for all $j$. The resulting *I-V* curve and *dV/dI* are shown in Fig. 3A of the main text.
It is important to note, that if $v_d$ approaches $v_F$, the Fermi energy $E_F$ calculated from Eq. (S7) decreases and can become smaller than $T$. In this case, temperature smearing of the Fermi surface is expected to introduce additional electrons and holes. Integrating the shifted Fermi distribution (S6) over the conduction and valance bands one can get

$$n_{e/h} = (\pi\hbar)^{-2} \int d^2 p\, n_F(\epsilon_p - p_x v_d \pm E_F, T) = -\frac{2T^2 k_B^2 Li_2(-e^{\pm E_F/T})}{\pi\hbar^2 v_F^2 (1 - v_d^2/v_F^2)^{3/2}} \tag{S8}$$

where $Li$ is the polylogarithm function, $n_e$ and $n_h$ are the electron and hole densities, respectively. Also, the total charge density $n$ controlled by electrostatic gating should be conserved as

$$n = n_e - n_h \tag{S9}$$

From equations (S8) and (S9) we can find the Fermi energy and the number of additional carriers created by temperature in the case of a strongly shifted Fermi surface. This effect becomes stronger with increasing $T$. It smoothens the critical transition and suppresses the peak in $dV/dI$ at elevated temperatures as seen in fig. S4.

#6 Generating electron-hole plasma by a Schwinger-like mechanism

In this section we describe $I$-$V$ characteristics at supercritical currents exceeding $j_c$. For simplicity, we assume a one-dimensional model such that electrons and holes propagate parallel or antiparallel to the applied electric field $E$. Assuming a diffusive flow, the current can be written as

$$j = ev_F(n_e + n_h) \tag{S10}$$

where $n_e$ and $n_h$ are the total concentrations of electrons and holes, respectively, including the carriers induced by gate voltage ($n$) and e-h pairs generated in interband transitions ($\Delta n$). Because all the carriers in our 1D model propagate along $E$, the drift velocity $v_d$ should be equal to $v_F$. Any increase in $j$ beyond $j_c$ can only happen due to extra carriers that are added to the system. To calculate $I$-$V$ characteristics, we therefore need to understand how $\Delta n$ changes with bias.

The equilibrium concentration of e-h pairs ($\Delta n$) is governed by a balance between creation and annihilation of extra carriers. Their creation rate in graphene is described by interband tunneling as $AE^{3/2}$, according to the Schwinger or Zener-Klein mechanism where $A$ is a dimensional constant (*16*). On the other hand, newly created carriers can also annihilate through e-h recombination at a rate of $Bn_e n_h$ (*5*) where $B$ is another dimensional constant. Changes in the carrier density can then be written as

$$\frac{dn_e}{dt} = AE^{3/2} - Bn_e n_h \tag{S11}$$

$$\frac{dn_h}{dt} = AE^{3/2} - Bn_e n_h \tag{S12}$$

Here we assume $n_{e/h}$ are spatially uniform and exclude any net flux due to charge-density gradients. For the steady-state regime ($\frac{dn_e}{dt} = \frac{dn_h}{dt} = 0$), equations (S11) and (S12) are reduced to

$$AE^{3/2} = Bn_e n_h \tag{S13}$$

Let us consider two opposite limits, $\frac{j-j_c}{j_c} \ll 1$ and $\frac{j-j_c}{j_c} \gg 1$. Without loss of generality, we assume that graphene is initially doped with electrons and write

$$\begin{cases} n_e = n + \Delta n \\ n_h = \Delta n \end{cases} \tag{S14}$$

Considering that $j_c \approx nev_F$ (see the main text) and using equations (S8) and (S13), we find that the limit $\frac{j-j_c}{j_c} \ll 1$ corresponds to $\Delta n \ll n$. In this limit, $E$ generates a relatively small concentration of e-h pairs so that we can use $n_e \approx n = $ const, whilst $E$ strongly alters the hole concentration, $n_h = \Delta n$. By placing these $n_e$ and $n_h$ into equation (S13), we obtain

$$\Delta n = \frac{A}{B} n^{-1} E^{3/2} \propto E^{3/2} \tag{S15}$$

Equations (S10), (S14) and (S15) yield the following current-voltage relation: $\Delta j = j - j_c \approx 2\Delta n ev_F \propto E^{3/2} \propto V^{3/2}$ where $V$ is the voltage drop along our channels. Rearranging this equation as $V \propto \Delta j^{2/3}$ and taking the derivative, we find the differential resistivity in this limit as

$$\frac{dV}{d\Delta j} \propto \Delta j^{-1/3} \tag{S16}$$

The negative power in equation (S16) comes from the fact that at supercritical currents the electric field starts generating new charge carriers at a rate $E^\gamma$ with $\gamma > 1$, which results in the differential resistivity decreasing with increasing $j$.

In the opposite limit of very high currents, $\frac{j-j_c}{j_c} \gg 1$, we obtain $\Delta n \gg n$, as expected. This is the regime of a compensated e-h plasma, $n_h \approx n_e \approx \Delta n$. Using equations (S13) and (S10), we then find $\Delta n = \frac{A}{B} E^{3/4} \propto E^{3/4}$ and $j = 2\Delta n e v_F \propto E^{3/4} \propto V^{3/4}$ or $V \propto j^{4/3}$. Taking the derivative, we obtain

$$\frac{dV}{dj} \propto j^{1/3} \tag{S17}$$

This equation shows that, in the high-bias regime such that $n_h \approx n_e \approx \Delta n$, the differential resistance should increase with increasing $j$. If graphene is doped with $n > 10^{11}$ cm$^{-2}$, it is hard to achieve this regime because of very large $j$ required to reach $\Delta n \gg n$. Such $j$ normally damage graphene devices. However, close to the NP, $\Delta n$ needs only to exceed the charge inhomogeneity level and, therefore, much smaller $j$ are sufficient to create a compensated e-h plasma, as seen in Fig. 3D of the main text.

#7 Critical drift velocity in graphene superlattices
To gain more information about the critical-current behavior in our superlattice devices, we evaluated drift velocities at which the switching transition occurred. Similar to our analysis for non-superlatticed graphene in the main text, we used the expression $v_d = j_c/n_S e$ where $n_S$ is the carrier density within different minibands. It is different from the total doping $n$ induced by gate voltage (Figs. 1A-C of the main text). To find $n_S$, we followed the procedure described in (*42*). Briefly, we first used Hall measurements in small magnetic fields (section #2) to extract the geometrical capacitance $C_g$ to the back gate and to find gate voltages $V_g$ for NPs and VHS ($V_{NP}$ and $V_{VHS}$, respectively). Next, assuming that different electronic minibands did not overlap so that $n_S$ should be zero at NPs, we calculated $n_S$ around them as $n_S = C_g(V_g - V_{NP})/e$. The resulting linear dependences $n_S(V_g)$ were separated by VHS where $n_S$ abruptly changed its sign (*42*).

The found $v_d$ are shown in figs. S5A,B for two TBG and three G/hBN devices. In contrast to the behavior in non-superlatticed graphene, where $v_d$ at the critical current was close to $v_F$ and independent of $n$ (Fig. 3B), superlattices exhibited profound changes in $v_d$ as a function of band filling. The drift velocity was highest near NPs, rapidly decreased away from them and became smallest around VHS. For the TBG 1.23° device, the observed $v_d$ close to the main NP was about 4-5×10$^4$ m s$^{-1}$, in good agreement with the Fermi velocity of 3×10$^4$ m s$^{-1}$ which we found from band structure calculations for this twist angle. Note that elastic reconstructions of the superlattice structure at this $\theta$ are expected to be small (unlike for $\theta \approx 0.77°$), which makes the calculations reliable (*19*). The agreement indicate that the switching transition in graphene superlattices is due to the same mechanisms as suggested for monolayer graphene (Fig. 3), except for the fact that its observation requires smaller $j$ because of lower Fermi velocities in the superlattices. This also explains the absence of the transition around the main NP in G/hBN superlattices where the Fermi velocity $\sim 10^6$ m s$^{-1}$ is more than 10 times higher than that near secondary NPs (fig. S5) and, accordingly, the required critical currents were not reachable near the main NP in the standard Hall bar geometry.

For all the studied superlattice devices, we found $v_d$ to rapidly decrease away from NPs. To explain this very general observation, we note that the group velocity of charge carriers changes with the Fermi energy. It is normally highest near NPs and drops to zero at VHS as schematically shown in fig. S5C. Only for the linear part of superlattice spectra the drift velocity is expected to coincide with $v_F$. Indeed, as the nonequilibrium distribution shifts away from a NP, electronic states near a neighboring VHS start contributing to the charge flow (fig. S5C). Those states have low group velocities, approaching zero at VHS, which leads to reduction in the average group velocity. This consideration qualitatively explains the experimental behavior in figs. S5A,B. Nonetheless, let us notice that the effect of nonlinear electronic spectra becomes pronounced already very close to NPs (that is, far away from VHS). It requires numerical simulations of 2D quantum transport for specific electronic spectra to explain this rapid decrease in $v_d$ quantitatively. The curves in Fig. S5 or maps in Fig. 2 of the main text could then be used to reconstruct superlattice spectra.

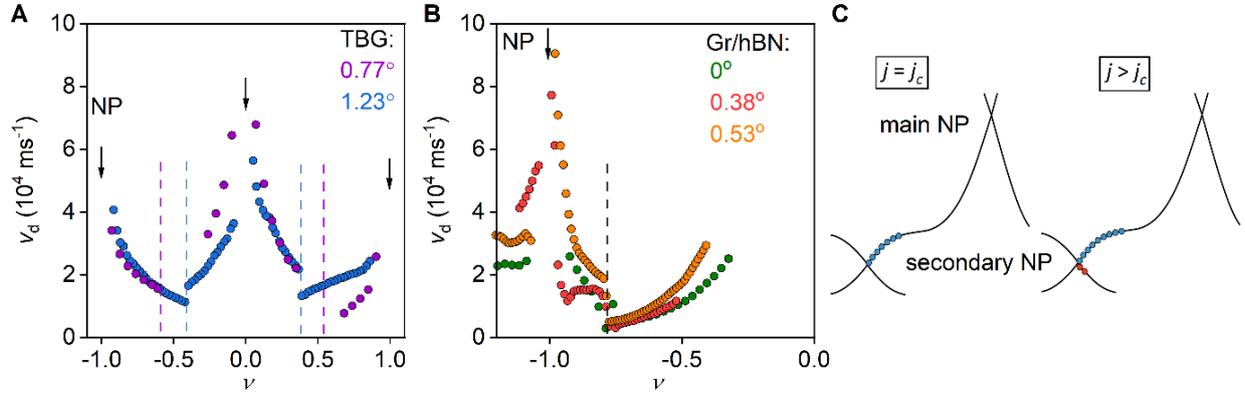

**Fig. S5. Drift velocities at the critical current values in various graphene superlattices.** (**A**) *Drift velocity $v_d$ at the peak in differential resistivity for our TBG devices. To show different superlattices on the same graph, we use the filling factor ν of the first miniband as the common x-axis. The black arrows mark NPs and the vertical dashed lines VHS. For TBG 0.77°, peaks in dV/dI were poorly resolved, and we defined $j_c$ using Hall effect measurements.* (**B**) *Same as (A) for three G/hBN superlattices (also, see fig. S6).* (**C**) *Schematics of the out-of-equilibrium carrier distribution in the valence band of G/hBN superlattices for $j = j_c$ (left) and $j > j_c$ (right). The blue circles show electronic states occupied by electrons and the red ones those occupied by holes.*

#8 Reproducibility for different G/hBN devices
Further examples of the switching transition between Fermi-liquid and e-h plasma transport regimes are shown in fig. S6 for G/hBN superlattices with non-zero twist angles $\theta \approx 0.38°$ and $0.53°$ (nonperfect alignment between graphene and hBN axes). The low-bias resistivities show the behavior standard for graphene superlattices (figs. S6A,D). Moving into the high-bias regime, both devices exhibited the critical behavior qualitatively similar to that reported for G/hBN 0° in the main text. Indeed, as seen in figs. S6B,E, the small-$\theta$ superlattices showed same switch-like increases from low to high resistance and pronounced peaks in *dV/dI*. The differential resistivity maps were also similar (figs. S6C,F) with the critical-current transition moving to progressively lower *j* as *n* approached the secondary NPs. The peak in *dV/dI* was progressively smeared and gradually disappeared towards lower hole concentrations, in agreement with the map shown in Fig. 2A of the main text.

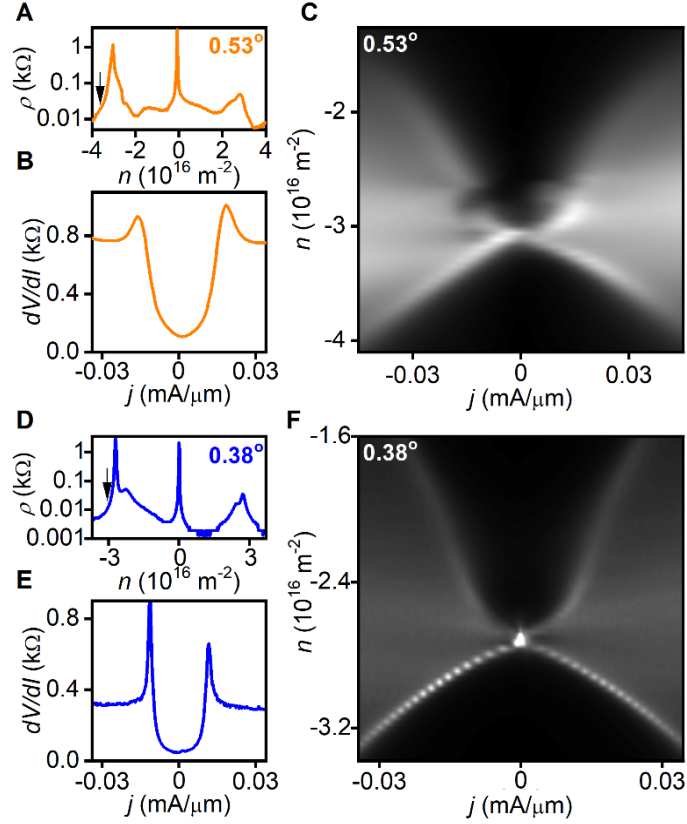

**Fig. S6. Critical-current behavior in other G/hBN superlattices.** (**A**) *Low-bias resistivity for a G/hBN superlattice with θ ≈ 0.53º.* (**B**) *Its differential resistivity dV/dI for the doping indicated by the black arrow in (A).* (**C**) *Map of the differential resistivity as a function of n and j. Grey scale: black-to-white, 10 to 2000 Ω. The transition is somewhat distorted just above the secondary NP, which is attributed to either charge or twist-angle inhomogeneity.* (**D-F**) *Same as in panels (A-C) for another G/hBN superlattice with θ ≈ 0.38º.*

#9 Nonlinear dual-band Hall effect
The generation of e-h pairs above the critical current $j_c$ resulted in the nonlinear Hall response as reported in the main text (Fig. 2). Here we show that the observed behavior can qualitatively be describes by the standard two-carrier model. In small (non-quantizing) magnetic fields $B$, the Hall voltage $V_{xy}$ is given by

$$V_{xy} = R_{xy}[I]I \qquad (S18)$$

$$R_{xy} = \frac{\sigma_{xy}}{\sigma_{xy}^2 + \sigma_{xx}^2} \qquad (S19)$$

$$\sigma_{xx} = \frac{en_e[I]\mu_e}{1+(\mu_e B)^2} + \frac{en_h[I]\mu_h}{1+(\mu_h B)^2} \qquad (S20)$$

$$\sigma_{xy} = \frac{en_h[I]\mu_h^2 B}{1+(\mu_h B)^2} - \frac{en_e[I]\mu_e^2 B}{1+(\mu_e B)^2} \qquad (S21)$$

where $\mu_e$ and $\mu_h$ are the electron and hole mobilities, respectively, and $I$ is the applied current. Nonlinearities in this model arise because of extra charge carriers generated above the critical current $I_c$ (corresponding to the critical current density $j_c$). To model the Hall effect as a function of $I$, we assume $n_{e/h}$ to be constant below $I_c$ whereas, above $I_c$, all the additional current ($\Delta I = I - I_c$) is carried by the newly created electrons and holes, both having densities $\Delta n$. This is formally described as

$$n_h = n_{h0} + \Delta n$$

$$n_e = n_{e0} + \Delta n$$

$$\Delta n = \begin{cases} 0, & I < I_c \\ \approx \frac{(I-I_c)}{2Wev_F}, & I > I_c \end{cases} \quad (S22)$$

where $n_{e0}$ and $n_{h0}$ are the carrier densities of electrons and holes in the linear response regime, and $v_F$ is the Fermi velocity near relevant NPs where the generation of new charge carriers takes place. For simplicity, we again assume transport to be 1D such that carriers propagate only along the electric field and their drift velocities are equal to $v_F$ for $I > I_c$. Note that, in the 2D case, the average drift velocity of the generated charge carriers is expected to be less than $v_F$ and, therefore, $\Delta n$ should be somewhat larger than that given by equation (S22).

An example of the nonlinear Hall behavior was provided in Fig. 2D of the main text and is now replotted in fig. S7 (solid red curve). The dashed curves in the same figure show three different fits using equations (S18-S22). The fitting parameter required to describe the unusual shape of experimental curves is the ratio between electron and hole mobilities ($\mu_e/\mu_h$), and only the amplitude of Hall signals depends on absolute values of $\mu_e$ and $\mu_h$. If newly created electrons and holes had same $\mu$, our modelling shows that Hall curves could not exhibit sharp changes with increasing $I$ above $I_c$. However, for strongly different $\mu_e$ and $\mu_h$, I-V characteristics start exhibiting clear turning points at $I_c$, in agreement with the experimental behavior (fig. S7). This can be understood as follows. The low-bias Hall signal produced by initial (low-mobility) holes near the VHS (see Fig. 2) becomes rapidly compensated by a contribution from high-mobility electrons generated by interband tunnelling near the secondary NP. With further increase of $I$, more and more high mobility electrons are added to the system, and they eventually dominate the Hall effect causing its sign reversal.

The inferred large difference in $\mu$ for newly created electrons and holes is expected because of the nonlinear spectrum of graphene superlattices near the secondary NPs, as already discussed in the main text and section #7. The difference could be understood using the sketch in fig. S5C, which represents the case of initial electron doping with the Fermi level lying between the secondary NP and a VHS (opposite initial doping to that in fig. S7). In the case of fig. S5C, interband transitions generate holes near the NP. These holes appear in the region of the linear spectrum and should have high mobilities, like generally happens in graphene-based systems with Dirac spectra. On the other hand, we expect low mobility for extra electrons. As seen in the sketch, the original electron distribution gets shifted towards higher energies for $I > I_c$, and extra electrons effectively appear close to the VHS and have high masses. Therefore, the electron mobility averaged over all electrons contributing into the current should be lower than $\mu$ for initial electrons near the NP and the generated holes.

#10 Further examples of I-V characteristics at the Dirac point in graphene

When studying I-V characteristics at the NP, it is essential to account for so-called self-gating. This phenomenon refers to the situation where a voltage drop induced by applied current becomes so large that it notably changes local carrier concentrations along the device. Therefore, at high biases, $n$ is no longer defined only by applied gate voltage but also depends on a voltage drop across the device. The main contribution into self-gating for our constrictions came from a voltage drop in the contact regions whereas the constriction itself experienced a small voltage drop (< 1V) that resulted in a relatively small charge inhomogeneity inside the constriction. Therefore, the main effect of high bias was a shift of $n$ within the constriction with respect to initial doping. To properly evaluate differential resistivity at the NP, this shift has to be carefully monitored, especially in high-quality devices with very sharp peaks in low-bias resistivity. To this end, rather than sweeping $j$ at a fixed back-gate voltage $V_g$, we measured $dV/dI$ at fixed $j$ and varied $V_g$. Then, $j$ was increased in small steps as illustrated in fig. S8A. One can see that the NP - normally seen as a resistance maximum - shifted with increasing $j$. The direction of this shift changed with reversing the applied DC current, as expected. By tracing NP positions, we reconstructed the $dV/dI$ dependence at the NP. Fig. 3D of the main text shows one set of such measurements, and further examples are provided in fig. S8B. One can see that all the curves exhibited critical behavior similar to that discussed in the main text.

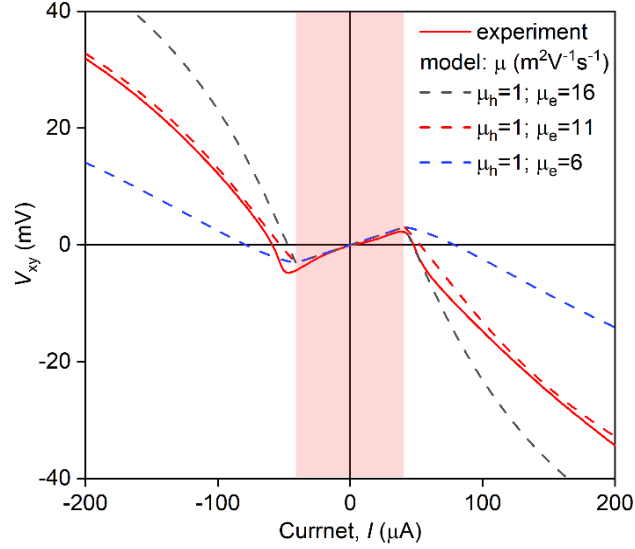

**Fig. S7. Modelling of the dual-band Hall effect.** *The solid curve is the Hall voltage measured at 30 mT (same as in Fig. 2D of the main text). Dashed curves: modelling using the mobilities shown in the color-coded legend. B = 30 mT; W = 4 μm; $I_c$ = 41 μA; $n_{h0} = 0.25 \times 10^{16}$ $m^{-2}$; $n_{e0} = 0$; $v_F = 0.4 \times 10^6$ m s$^{-1}$. The latter is the Fermi velocity expected at the secondary NP of G/hBN superlattices (46). The shadow region indicates the regime $|I| < I_c$.*

#11 Quantum-critical resistivity of the Dirac fluid

To evaluate resistivity $\rho$ of the hot e-h plasma from our measurements on graphene constrictions at the Dirac point, we need to take into account that the resistance $R$ measured in the quasi-4-probe geometry involves contributions of not only the constriction itself but also nearby regions leading to voltage contacts (inset of Fig. 3A of the main text). First, it is instructive to consider this geometry in the linear response regime where graphene's resistivity $\rho$ is independent of $j$. Approximating the areas near the constriction as semicircles, their contribution can be written as $2 \int_{W/2}^{H} \rho(j=0) \frac{dr}{\pi r}$ where $r$ is the distance from the constriction center, $W$ is the constriction width, and the factor of 2 accounts for the areas left and right from the constriction. Here $H$ is the cutoff at long distances which depends on geometry and determines the effective area contributing to the measured $R$. The constriction has a physical length $L$ but, considering that the same current density persists over an area of size $\sim W$ around the constriction (see below), we approximate the central channel as a square, that is, the effective length $L = W/2$ from the constriction center. The total resistance of the constriction can then be written as

$$R = 2 \int_{W/2}^{H} \rho \frac{dr}{\pi r} + \rho \frac{2W/2}{W} = \rho \left( \frac{2}{\pi} \ln \frac{2H}{W} + 1 \right) \tag{S23}$$

The lower cutoff in the integral is given by $L$ whereas the higher cutoff $H$ is of the order of devices' width away from the constriction, which in our case was $\sim 10W$. This imposes the following upper and lower bounds on $R$

$$1.4\rho < R < 2.9\rho \tag{S24}$$

which translates into $\rho$ of about 2.5 kΩ (± 50%) for our typical $R \approx 5$ kΩ in fig. S8B.

In the supercritical regime, a more careful estimate for the relation between $R$ and $\rho$ is required because the differential resistance becomes a function of $j$ (Fig. 3D and fig. S8) and, therefore, a function of $r$. To account for these dependences, we use Eq. (S17) and model $\rho(j)$ as $\propto j^{1/3}$ or

$$\rho(r) = \rho_0 \left(\frac{j(r)}{j_0}\right)^{1/3} \tag{S25}$$

where $\rho_0$ and $j_0$ are some dimensional constants. In the 2D case, the current density decreases away from the constriction as

$$j(r) = \frac{I}{\pi r} \tag{S26}$$

The current density inside the constriction is given by $j_{MAX} = I/W$. According to Eq. (S26), $j$ reaches the same maximum value at $r = W/\pi$. The latter distance serves as a cutoff for the applicability of the above formula at short $r$. We therefore assume the constant current density $j = j_{MAX}$ within the central area of the constriction, which has the width $W$ and half-length $W/\pi$, whereas outside of this rectangle $j$ decays as prescribed by Eq. (S25). We are interested in finding the hot e-h plasma's resistivity inside the constriction $\rho(r = 0)$, which within our formalism is the same as $\rho(r = W/\pi)$.

As for a cutoff at long distances, we note that the e-h plasma fully develops only for $j > j_m$ (see the main text). This condition yields the size $H$ of the area in which the high-resistivity e-h plasma is created as

$$\frac{I}{\pi H} = j_m \text{ and, hence, } H = \frac{W j_{MAX}}{\pi j_m} \tag{S27}$$

In our experiments, typical $j_m \approx 0.05$ mA/μm whereas $j_{MAX}$ reaches 0.7 mA/μm (Fig. 3D and fig. S8). This means $H \approx 4.5W$. Using the above cutoffs at short and long distances, we evaluate the high-bias resistance for our constrictions as

$$R = 2 \int_{W/\pi}^{H} \rho(r) \frac{dr}{\pi r} + \rho(r = W/\pi) 2/\pi \tag{S28}$$

where the second term is the approximation for the resistance of the central rectangular region. The integration yields

$$R \approx 1.76 \rho(r = 0) \tag{S29}$$

For our maximum applied $j$, the resistance measured across the constrictions was ~4.5 – 5 kΩ (fig. S8) which yields resistivity of the hot e-h plasma as $\rho \approx 2.7$ kΩ (±10%). The fact that this value is close to that found in the low-bias regime indicates that geometrical factors in translating $\rho$ into $R$ depend only logarithmically on current distribution, which ensures the reliability of our analysis.

Finally, we compare the found $\rho$ of the compensated e-h plasma with theory (8, 9, 41, 43). The scattering rate $\nu$ between electrons and holes in such an e-h plasma (Dirac fluid) is expected to approach the quantum critical limit

$$\nu = C \frac{k_B T}{\hbar} \tag{S30}$$

where $C$ is the numerical constant of the order of 1. As the carrier density in the Dirac fluid is also proportional to its temperature $T$, one finds the quantum critical resistivity (43)

$$\rho_q = \frac{h}{e^2}\frac{1}{2\ln 2}\frac{\hbar v}{k_B T} = \frac{h}{e^2}\frac{C}{2\ln 2} = C \times 18.6\ (k\Omega) \qquad (S31)$$

The value of ~2.7 kΩ found above is in reasonable agreement with a recent experiment (*11*) that suggested $C \approx 0.2$ for the Dirac fluid and, therefore, eq. (S31) would yield $\rho_q \approx 3.7\ k\Omega$. More accurate theory analyses of $\rho_q$ have found (*8*, *9*)

$$\rho_q = \frac{h}{e^2}\frac{\alpha^2}{0.764} \qquad (S32)$$

where $\alpha$ is the interaction constant that, importantly, depends on the dielectric constant $\varepsilon$ of the media surrounding graphene. Accounting for screening (*8*), $\alpha = \alpha_0/(1 + \pi\alpha_0/2)$ where $\alpha_0 \approx 2.1/\varepsilon$ is the effective fine structure constant in graphene. As $\varepsilon$ of encapsulating hBN is ~4.0, we obtain $\alpha \approx 0.3$ and, accordingly, $\rho_q \approx 3\ k\Omega$. This is in excellent agreement with our experiment yielding $\rho_q \approx 2.7\ k\Omega$, especially because the curves in fig. S8 continue to saturate with increasing *j* to a somewhat higher value.

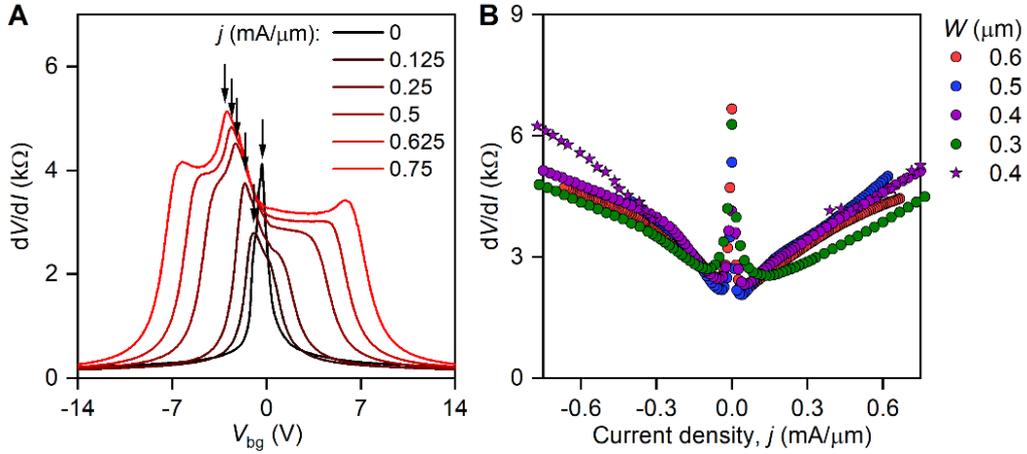

**Fig. S8. High-bias response at the NP in non-superlatticed graphene.** (**A**) *Differential resistance as a function of back-gate voltage for different current densities inside a 0.4 µm wide constriction (color coded). The arrows mark positions of the NP for different j.* (**B**) *Differential resistance at the NP measured for several graphene constrictions.*

#12 Calculations of superlattice spectra
To calculate the electronic spectrum of G/hBN superlattice in Fig. 1D of the main text, we used the technique described in (*18*) where, for simplicity, only the inversion-symmetric coupling parameters were included [see Eq. 2 of (*18*)]. Following (*44*), we set the parameters as $u_0 v_F b = 8.5\ meV$, $u_1 v_F b = -17\ meV$, $u_3 v_F b = -14.7\ meV$.
For the spectrum of TBG shown also in Fig. 1D, we used the approach described in (*19*) with the hopping energy set to $w = 110\ meV$. A *Mathematica* code for these calculations is provided in (*45*).